\documentclass[french,british,preprint, aps,prl]{revtex4-2}
\usepackage[T1]{fontenc}
\usepackage[latin9]{inputenc}
\usepackage{color}
\usepackage{bm}
\usepackage{amsmath}
\usepackage{amssymb}
\usepackage{graphicx}

\makeatletter

\pdfpageheight\paperheight
\pdfpagewidth\paperwidth

\providecommand{\tabularnewline}{\\}

\newcommand{\rr}{\bm{r}}

\makeatother

\usepackage{babel}
\makeatletter
\addto\extrasfrench{%
   \providecommand{\fg}{\ifdim\lastskip>\z@\unskip\fi~\frqq}%
}

\makeatother
\begin{document}
\title{A Molecular Density Functional Theory of aqueous electrolytic solution.}
\author{\selectlanguage{french}%
Guillaume Jeanmairet}
\affiliation{\selectlanguage{french}%
Sorbonne Université, CNRS, Physico-Chimie des Électrolytes et Nanosystèmes
Interfaciaux, PHENIX, F-75005 Paris, France}
\affiliation{\selectlanguage{french}%
Réseau sur le Stockage Électrochimique de l\textquoteright Énergie
(RS2E), FR CNRS 3459, 80039 Amiens Cedex, France}
\author{\selectlanguage{french}%
Luc Belloni}
\affiliation{\selectlanguage{french}%
LIONS, NIMBE, CEA, CNRS, Université Paris-Saclay, 91191 Gif-sur-Yvette,
France}
\author{\selectlanguage{french}%
Daniel Borgis}
\affiliation{\selectlanguage{french}%
PASTEUR, Département de chimie, École normale supérieure, PSL University,
Sorbonne Université, CNRS, 75005 Paris, France}
\affiliation{\selectlanguage{french}%
Maison de la Simulation, CEA, CNRS, Universit\'{e} Paris-Sud, UVSQ,
Universit\'{e} Paris-Saclay, 91191 Gif-sur-Yvette, France}
\selectlanguage{british}%
\begin{abstract}
We propose a generalisation of molecular density functional theory
to describe inhomogeneous solvent mixture, with the objective of modelling
electrolytic solutions. Two electrolytic models are presented, both
within the HNC approximation. The first one is a two-components mixture
representing a primitive-like model of sodium chloride, where the
solvent is described as a dielectric continuum. This popular model
has the advantage of simplicity, as the ions densities solely depend
on spatial coordinates. Additionally, we develop a realistic three-components
electrolyte model, in which water solvent is described by a third
density field that depends on both spatial and orientational coordinates.
The proposed methodology and its tridimensional implementation (3
spatial coordinates and 3 Euler angles) are validated by comparing
the solvation properties of a sodium cation with the predictions of
integral equation theory solved in 1D (1 intermolecular distance and
5 Euler angles), showing near-perfect agreement. This methodology
enables the study of solvation properties of solutes of arbitrary
shapes in electrolytic solutions, as demonstrated with the prototypical
N-methylacetamide molecule immersed in both electrolytic solution
models.
\end{abstract}
\maketitle

\section{Introduction}

Electrolytic solutions, i.e solutions containing ions, are present
in a wide variety of chemical systems. For instance, ions play a key
role in biology, influencing protein stability\citep{broering_evaluation_2005}
and protein-protein interactions\citep{dumetz_patterns_2007}. Another
economically significant application is electrochemical storage, particularly
Li-ion batteries and supercapacitors. While Li-ion batteries store
energy through Faradaic processes, supercapacitors store charge via
ion adsorption at electrode surfaces. The interfacial region, where
ion concentrations deviate from equilibrium, is termed the electrical
double layer (EDL).

Therefore, in computer simulations of biological objects and electrochemical
devices, it is essential to have a realistic description of the electrolyte,
especially of the electrostatic interactions. The most accurate strategy
is to explicitly describe the constituents of the electrolyte. This
is the strategy adopted in molecular simulations such a Monte Carlo
and molecular dynamics (MD), which has been successfully employed
for the modelling of many biological objects, ranging form proteins\citep{friedman_molecular_2005}
to DNA\citep{xue_transformation_2022}, and for supercapacitors\citep{merlet_molecular_2012,pean_dynamics_2014,salanne_efficient_2016}.
However, these approaches become very costly when simulating large
systems and/or when long simulation times are required, partly because
of the large number of solvent and electrolyte molecules required
to solvate the system. 

Following the pioneering work of Gouy, Chapman, and Stern (GCS)\citep{kornyshev_double-layer_2007},
there is a long tradition of investigating electrolyte at interface
using theoretical approaches rather than simulations. A common strategy
is to use implicit solvent models in which the solvent, usually water,
is described by a dielectric continuum. In practice, the idea is to
define a cavity where lies the solute, for instance a protein. The
dielectric constant is assumed to vary from its bulk value outside
the cavity towards another small value inside the protein. The value
of the dielectric inside the protein\citep{fogolari_poissonboltzmann_2002,baran_electrostatic_2008}
and the shape of the boundary\citep{connolly_computation_1985,nina_optimized_1999}
influences the predictions, which explain why several choices have
been proposed for both. The electrostatic potential can then be obtained
by solving the Poisson equation. It is also possible to include ions
at the same level of description, whose density is assumed to follow
Boltzmann statistics. The solution of the Poisson equation for the
electrostatic potential coupled with the Boltzmann statistics for
ions give rise to the popular Poisson-Boltzmann (PB) model of electrolytes. 

This model suffers from several limitations, in particular the ions
are assumed to be point charges. The absence of size effects can lead
to abnormally large ionic concentrations close to charged solutes.
Another limitation is that ion-solvent and ion-ion correlations are
neglected. Several modification of the Poisson-Boltzmann model have
been proposed to address these flaws\citep{gavryushov_electrostatics_2008,grochowski_continuum_2008,ren_biomolecular_2012}. 

There is another strategy that avoids the tedious sampling of the
solvent degrees of freedom while maintaining a microscopic description
of the solvent, that is to use liquid state theories. The main strategy
is to use integral equation theory (IET) in which the components of
the solvent are no longer described by a set of discrete particles.
Instead, the average probability to find the solvent components at
a given position with respect to the solute is contained into some
distribution functions. By solving a coupled set of integral equations,
namely the exact Ornstein Zernike (OZ) equation, associated with an
approximated closure relation, it is possible to have access to the
solvent structure surrounding the solute. 

However,\textcolor{blue}{{} }\textcolor{black}{solving the OZ equation
for molecular solvents, such as water, and complex 3D solutes, as
encountered in biology, can be difficult due to the high number of
dimensions. This is why further approximations were proposed}, a popular
one being the reference interaction site model (RISM)\citep{chandler_optimized_1972,hirata_extended_1981}
which approximates the molecular solvent distribution by distributions
of atomic sites. This approach, and its 3D-RISM version, has been
widely employed to study small molecular solutes, either described
with classical force fields or quantum mechanics\citep{beglov_integral_1997,kovalenko_three-dimensional_1998,kovalenko_potentials_2000,kloss_quantum_2008,hoffgaard_three-dimensional_2013}. 

IET has also been adopted to simulate proteins\citep{imai_water_2005,yoshida_molecular_2009,sugita_new_2020,osaki_3d-rism-ai_2022}
and DNA\citep{maruyama_revisiting_2010,giambasu_competitive_2015}.
Despite its success, the 3D-RISM theory is not physically well-grounded,
in particular, the assumption that the molecular direct correlation
function can be decomposed as a sum of site-site correlation function\citep{hirata_molecular_2003},
which is at the core of the theory, is not valid and relies on a crude
approximation of the intramolecular correlations. 

\textcolor{black}{Classical density functional theory (cDFT) is a
functional formulation which is strictly equivalent to IET. Th}e principles
of cDFT are the following\citep{mermin_thermal_1965,evans_nature_1979}:
i) There exists a unique functional of the particle densities. ii)
This functional is equal to the system's grand potential at its minimum
which iii) is reached for the equilibrium particle densities. Therefore,
structural and thermodynamic properties can be obtained through functional
minimisation\textcolor{black}{{} rather than by solving integral equations,
these latter expressing that the gradient of the functional is zero.}

In cDFT, the ionic components of the electrolyte are modelled by their
spatially dependant density fields, $n_{+}$ and $n_{-}$ with the
possibility to explicitly account for the solvent with a third density
field, $n_{\text{s}}$. When a perturbation is present, the particles
densities become inhomogeneous. Due to this computational efficiency,
cDFT has been used to study electrolytic solution since its early
ages\citep{groot_density-functional_1988,mieryteran_nonlocal_1990,kierlik_density-functional_1991}.
Initial attention focused on the primitive model of electrolyte, consisting
of oppositely charged hard-sphere in a dielectric continuum. 

One of the key achievements of classical DFT is fundamental measure
theory (FMT)\citep{rosenfeld_free-energy_1989,kierlik_density-functional_1991,roth_fundamental_2002,hansen-goos_density_2006}
which provides nearly exact results for the hard-sphere fluid. As
a result, most recent works use FMT as their starting point to build
a functional for electrolytes. In contrast, the description of electrostatic
correlations is less straightforward, leading to the development of
various strategies. The simplest approach is to treat electrostatic
correlation at the mean-field level. When only this correlation is
considered, the theory is equivalent to Poisson-Boltzmann theory\citep{forsman_classical_2015}.
Incorporating finite-size effects and hard-sphere correlations makes
the theory akin to modified Poisson-Boltzmann approaches\citep{azuara_incorporating_2008}.

To go beyond mean-field, a common strategy is to perform a Taylor
expansion of the functional, truncated at second order, around a reference
fluid\citep{hartel_fundamental_2015,cats_capacitance_2022,mieryteran_nonlocal_1990}.
However, it is known that this quadratic truncation introduces thermodynamic
inconsistencies and other electrostatic functionals based on the weighted-density
approximation (WDA)\citep{levin_electrostatic_2002,groot_ion_1991,roth_shells_2016}
have been proposed to correct this defect. 

There is currently a growing interest in the development
of machine-learning-based functionals (MLF)\citep{lin_analytical_2020,cats_machine-learning_2021,sammuller_neural_2023},
including the development of a MLF for the restricted primitive model\citep{bui_learning_2025}.

Regarding the description of the solvent, it may be modelled implicitly
through a dielectric continuum that screens the interactions\citep{groot_density-functional_1988,groot_ion_1991,cats_capacitance_2022,mieryteran_nonlocal_1990,yang_systematic_2015},
or explicitly as an additional species. Here again, due to the quality
of FMT, the solvent is often modelled as an hard-sphere\citep{oleksy_towards_2006}
or as combination of hard-spheres\citep{jiang_solvent_2012,jiang_microscopic_2013,lian_capacitive_2017},
possibly with added charges or dipolar moments\citep{biben_generic_1998,oleksy_wetting_2010,oleksy_wetting_2011}. 

In this paper, we introduce a cDFT framework with the objectives of
achieving a more realistic description of both the solvent and the
solute. To do so, we will show in section \ref{sec:Theory} how the
molecular density functional theory (MDFT)\citep{jeanmairet_molecular_2013-1}
framework can be extended to multi-components fluids, \textit{$i.e$}
mixtures. The particularities due to the description of charged species,
$i.e$\textcolor{black}{{} ions, and the modifications it implies for
numerical implementation of MDFT will be discussed on the particular
case of aqueous electrolytic solution. Here, the solution is described
as a three-component mixture made of anion, cation and water. The
main result of this paper is this functional which describes an electrolytic
solution in a molecular solvent. For comparison, a primitive like
two-component functional is also presented.}

\textcolor{black}{The 3D MDFT implementation is validated by comparison
with the reference 1D-IET computation in section \ref{sec:Results}.
This validation is done by comparing the prediction of a sodium cation
in a 1M NaCl aqueous solution, with the two electrolyte models. Following
validation, we showcase the th}eory's capability to analyse complex
solutes by examining the solvation of an N-Methylacetamide molecule
in the same electrolyte solution models. This molecule involves a
single peptide bond and is often considered as the simplest realistic
model for the peptide group\citep{beicastro_density_1996}.

\section{\label{sec:Theory}Theory}

\subsection{Molecular Density Functional Theory of Mixtures}

Let us consider a $N$-component mixture of molecular species. Each
molecule is assumed to be rigid such as the knowledge of the position
$\bm{r}$ of its center of mass (COM) and of its absolute orientation
$\bm{\Omega}$ is enough to fully describe its set of coordinates. 

The objective is to develop a cDFT, which is most naturally formulated
in the Grand Canonical ensemble where V, T and the chemical potentials
$\mu_{\text{A}}$, of all species $\text{A}$ are fixed. Since the
entities are molecular and rigid, their one particle densities depend
on the space and angular coordinates, $i.e$
\begin{equation}
\rho_{\text{A}}(\bm{r},\bm{\Omega})=\left\langle \sum_{i\in\text{A}}\delta(\bm{r}-\bm{r}_{i})\delta(\bm{\Omega}-\bm{\Omega}_{i})\right\rangle ,
\end{equation}
where $\rho_{\text{A}}$ is the density of species A, and the sum
runs over each molecule of this species. $\delta$ denotes the Dirac
distribution, and the angle brackets indicate the ensemble average
in the Grand Canonical ensemble. The angular dependent density $\rho_{\text{A}}$
is related to the usual number density $n_{\text{A}}$ through:
\begin{equation}
n_{\text{A}}(\bm{r})=\int\rho_{\text{A}}(\bm{r},\bm{\Omega})d\bm{\Omega}.
\end{equation}

In the presence of an external perturbation, such as in the vicinity
of a solute, the molecular densities become inhomogeneous. Following
a derivation similar to Mermin and Evans' \citep{mermin_thermal_1965,evans_nature_1979}
pioneering work or using Levy-Lieb constrained search\citep{dwandaru_variational_2011,jeanmairet_rigorous_2024},
it is possible to prove the existence of a unique functional of the
set of solvent densities, $\Omega[\{\rho_{1},\rho_{2}...\rho_{N}\}]$,
which reaches its minimum when each density equals its equilibrium
value. At this minimum, the functional is equal to the grand-potential.
Note that this is a straightforward generalisation of the original
one-component cDFT which has already been widely used in the literature,
in particular for hard-body fluids\citep{rosenfeld_free-energy_1989,kierlik_density-functional_1991,roth_fundamental_2002,hansen-goos_density_2006}.
The originality of the present work lies in developing functionals
capable of describing mixtures of molecular fluids. Following the
strategy we adopt for one-solvent MDFT, we work with a new functional
which is defined as the difference between the functional of the perturbed
fluid and the grand potential of the homogeneous mixture, $\Omega_{0}$

\begin{equation}
F=\Omega[\left\{ \rho_{1},\rho_{2}...\rho_{N}\right\} ]-\Omega_{0}.
\end{equation}

As usual\citep{evans_nature_1979,hansen_theory_2006}, the functional
is split into the sum of three components
\begin{equation}
F[\left\{ \rho_{1},\rho_{2}...\rho_{N}\right\} ]=F_{\text{id}}[\left\{ \rho_{1},\rho_{2}...\rho_{N}\right\} ]+F_{\text{ext}}[\left\{ \rho_{1},\rho_{2}...\rho_{N}\right\} ]+F_{\text{exc}}[\left\{ \rho_{1},\rho_{2}...\rho_{N}\right\} ].\label{eq:F=00003DFid+fExc+Fext}
\end{equation}

The ideal part, which is the first term on the right hand side of
equation \ref{eq:F=00003DFid+fExc+Fext}, measures the entropic cost
for the system to acquire non-homogeneous densities. In this context,
it corresponds to the functional of a mixture of non-interacting fluids.
\begin{equation}
F_{\text{id}}[\left\{ \rho_{1},\rho_{2}...\rho_{N}\right\} ]=k_\text{B}T\sum_{\text{A}=1}^{N}\iint\left[\rho_{\text{A}}(\bm{r},\bm{\Omega})\ln\left(\frac{\rho_{\text{A}}(\bm{r},\bm{\Omega})}{\rho_{0\text{A}}}\right)-\Delta\rho_{\text{A}}(\bm{r},\bm{\Omega})\right]d\bm{r}d\bm{\Omega}\label{eq:Fid}
\end{equation}
where the sum runs over the constituents of the mixture and $\Delta\rho_{\text{A}}=\rho_{\text{A}}-\rho_{0\text{A}}$
is the excess density of $\text{A}$ with respect to the homogeneous
density $\rho_{0\text{A}}$. The second term is due to the interaction
with the external field
\begin{equation}
F_{\text{ext}}[\left\{ \rho_{1},\rho_{2}...\rho_{N}\right\} ]=\sum_{\text{A}=1}^{N}\iint\rho_{\text{A}}(\bm{r},\bm{\Omega})V_{\text{ext}}^{\text{A}}(\bm{r},\bm{\Omega})d\bm{r}d\bm{\Omega}\label{eq:Fext}
\end{equation}
where $V_{\text{ext}}^{\text{A}}$ are external potentials describing
the interactions between the solute and the different components of
the mixture. In this paper, as in most of our work, the external potentials
are parametrised with classical force fields composed of point charges
and Lennard-Jones sites. As an alternative, we recently proposed a
QM/MDFT framework where the electrostatic part of the external potential
is derived from the electronic density, computed using a quantum mechanics
description of the solute\citep{jeanmairet_tackling_2020,labat_coupling_2024,jeanmairet_rigorous_2024,labat_prediction_2025}.

The last term of equation \ref{eq:F=00003DFid+fExc+Fext} is the excess
term, which arises from the interactions between the particles of
the solution. It is possible to perform a systematic expansion of
this functional around the homogeneous densities.
\begin{align}
F_{\text{exc}}[\left\{ \rho_{1},\rho_{2}...\rho_{N}\right\} ]= & -\frac{1}{2}k_\text{B}T\sum_{\text{A}=1}^{N}\sum_{\text{B}=1}^{N}\iint\left[\Delta\rho_{\text{A}}(\bm{r}_{1},\bm{\Omega}_{1})c_{\text{AB}}(\left\Vert \bm{r}_{1}-\bm{r}_{2}\right\Vert ,\bm{\Omega}_{1},\bm{\Omega}_{2})\right.\nonumber \\
 & \left.\Delta\rho_{\text{B}}(\bm{r}_{2},\bm{\Omega}_{2})\right]d\bm{r}_{1}d\bm{r}_{2}d\bm{\Omega}_{1}d\bm{\Omega}_{2}+F_{\text{Bridge}}[\left\{ \rho_{1},\rho_{2}...\rho_{N}\right\} ].\label{eq:Fexc}
\end{align}
In equation \ref{eq:Fexc}, $F_{\text{Bridge}}$ is the so-called
bridge term, which contains all terms of order higher than 2 in $\Delta\rho$.
We have previously worked on several model of bridge functional for
pure solvent\citep{levesque_solvation_2012,jeanmairet_molecular_2015,borgis_simple_2020,borgis_accurate_2021,hsu_assessing_2021},
but their transferability to mixture should be studied carefully.
Therefore, the bridge term of equation \ref{eq:Fexc} will be omitted
in this paper. \textcolor{black}{This corresponds to the solute-solvent
HNC approximation in IET language.} The second order term involves
the $c_{\text{AB}}$ direct correlation functions (dcf) between species
$\text{A}$ and species $\text{B}$ in the homogenous (unperturbed)
mixture. The dcf are functions of the relative position and orientations
1, 2 (one distance and five Euler angles). 

The dcf can be obtained by solving the molecular Ornstein Zernike
equation for the homogeneous mixture at a given temperature and bulk
density and are inputs of the present theory. The exact dcf are obtained
using Monte-Carlo data at short distances and complemented with the
hypernetted chain closure, which is known to be valid at long distances\citep{belloni_efficient_2014,belloni_exact_2017}. 

Minimisation of equation \ref{eq:F=00003DFid+fExc+Fext} leads to
the following HNC integral equations
\begin{equation}
\rho_{\text{A}}(\bm{r},\bm{\Omega})=\rho_{0\text{A}}\exp[-\beta V_{\text{ext}}^{\text{A}}(\bm{r},\bm{\Omega})+\gamma_{\text{A}}(\bm{r},\bm{\Omega})].\label{eq:rho=00003DExp}
\end{equation}
The indirect solute-solvent pair correlation function $\gamma_{\text{A}}$
is given by the Ornstein-Zernike (OZ) solute-solvent equation 
\begin{equation}
\gamma_{\text{A}}(\bm{r},\bm{\Omega})=\sum_{\text{B}=1}^{N}\iint c_{\text{AB}}(\left\Vert \bm{r}-\bm{r}_{2}\right\Vert ,\bm{\Omega},\bm{\Omega}_{2})\Delta\rho_{\text{B}}(\bm{r}_{2},\bm{\Omega}_{2})d\bm{r}_{2}d\bm{\Omega}_{2}.\label{eq:gamma}
\end{equation}

The excess functional of equation \ref{eq:Fexc} and the integral
equation \ref{eq:gamma} involve convolution products. 

For molecular solvent, such as water, the angular convolution calculation
benefits from the use of an expansion onto a basis of generalized
spherical harmonics (GSH)\citep{blum_invariant_1972,ding_efficient_2017}.
In Fourier space and in the local intermolecular frame, the integral
equation \ref{eq:gamma} becomes a product between different projections,
characterised by three indices for the solute-solvent density distributions,
and five indices for the solvent-solvent dcf:
\begin{equation}
\hat{\gamma}_{A\mu;\chi}^{\ \ m}(\bm{k})=\sum_{B=1}^{N}\sum_{n,\nu}\left(-1\right)^{\chi+\nu}\hat{c}_{AB\mu\nu;\chi}^{\ \ \ mn}(k)\Delta\hat{\rho}_{B-\nu;\chi}^{\ \ \ n}(\bm{k}).\label{eq:gamma en proj}
\end{equation}

Note that the exchange between direct and Fourier spaces is made through
Fourier-Hankel transforms of projections defined in the fixed, laboratory
frame. The $\chi$ projections, in the local intermolecular frame,
used in equation \ref{eq:gamma en proj} can be obtained using the
standard $\chi$-transform of  Blum\citep{blum_invariant_1972-1,blum_invariant_1972,blum_invariant_1973}.
The angular convolution of equation \ref{eq:gamma} has thus been
replaced by matrix products. Moreover, the different values of $\chi$
are not mixed; there is one simple matrix equation \ref{eq:gamma en proj}
for each value of $\chi$. While, in principle, the basis of GSH is
infinite, it is truncated for values of $m,n\leq m_{\text{max}}$
in practice. More details about the overall procedure to compute $\gamma_\text{A}$ can be found in reference \citep{ding_efficient_2017}.  For monoatomic species, there are no angular degrees
of freedom, thus $m_{\text{max}}=0$ and the unique projection of
the density identifies with the number density: $\rho_{0;0}^{0}(\bm{r})=n(\bm{r})$. 

The functional of equation \ref{eq:F=00003DFid+fExc+Fext} is minimised
numerically using the following procedure. The densities are computed
in a orthorhombic box with periodic boundary conditions. They are
discretised on a $N_{\text{x}}\times N_{\text{y}}\times N_{\text{z}}$
spatial grid. Regarding the orientation in the laboratory frame, the first Euler angle $\theta$
is discretised using a Gauss-Legendre quadrature while a regular discretisation
is used for the remaining angles $\phi\text{ and }\psi$. The number
of angles is related to the choice for $m_{\text{max}}$\citep{ding_efficient_2017}.

The cycle starts with a guess for the densities $\rho_{\text{A}}(\bm{r},\bm{\Omega})$,
from which the functional of equation \ref{eq:F=00003DFid+fExc+Fext}
and its gradient at each grid point are computed. We take advantage
of the fast Fourier transform (FFT), computed with the FFTW3 package\citep{frigo_design_2005}
to handle the spatial convolution, while the angular convolution is
taken care of through the use of projections on rotational invariants
using equation \ref{eq:gamma en proj} following the procedure described
for one component solvent\citep{ding_efficient_2017}. The quasi-Netwon
LBFGS optimiser\citep{zhu_algorithm_1997} is used to propose a set
of new densities $\rho_{\text{A}}(\bm{r},\bm{\Omega})$. This procedure
is iterated until convergence is reached. Note that due to the logarithm
in the expression of the ideal functional in equation \ref{eq:Fid},
negative values of the density must be prevented during the optimisation
process. This condition is enforced by using auxiliary functions $\xi_{\text{A}}$
as the minimisation variables. There are related to the density through
the relationship
\begin{equation}
\rho_{\text{A}}(\bm{r},\bm{\Omega})=\rho_{0\text{A}}\xi_{\text{A}}^{2}(\bm{r},\bm{\Omega})
\end{equation}

The presence of charged particles, such as ions, in the solvent
mixture requires special treatment. Due to the long-range $1/r$ Coulombic
potential, the ion-ion dcfs $\hat{c}_{\text{AB}}(q)$ diverge as $-q_{\text{A}}q_{\text{B}}/k^{2}$
at $k=0$, where $q_{\text{A}}$ is the charge of species $\text{A}$.
These divergences impose that the ionic density profiles must fulfil
the electroneutrality condition for the total cell, during the whole
optimisation process. In order to avoid numerical divergence, we somewhat
artificially impose electroneutrality by using an alternative definition
for the density of the 1:1 salt as:
\begin{equation}
n_{+}(\bm{r})=n_{0}V\left(1+\lambda-\frac{Q}{n_{0}V}\right)\frac{\xi_{+}^{^{2}}(\bm{r})}{\int\xi_{+}^{^{2}}(\bm{r})d\bm{r}}\label{eq:densityNa-1}
\end{equation}
\begin{equation}
n_{-}(\bm{r})=n_{0}V\left(1+\lambda\right)\frac{\xi_{-}^{^{2}}(\bm{r})}{\int\xi_{-}^{^{2}}(\bm{r})d\bm{r}}\label{eq:densityCl-1}
\end{equation}
where $Q$ is the total charge of the solute. With this definition,
the electroneutrality is fulfilled, as $\int\left(n_{+}(\bm{r})-n_{-}(\bm{r})\right)d\bm{r}+Q=0$.
The additional variable $\lambda$ allows the number of ions in the
cell to fluctuate. 

The present 3D version of MDFT, where the spatial positions of the
solvent molecules are described by the $(x,y,z)$ coordinates of their
center of mass, is able to manage multi-site solutes of any complex
geometry. As a validation test for the proposed methodology, the solvation
of simple spherical ions which are constituents of the 1:1 electrolyte
will be considered below. In that case, the full 3D-methodology is
not optimal since obvious symmetry relations reduce the number of
independent parameters. For such solutes, it is preferable to use
a 1D version, where the density profiles depend on the distance $r$
between the solute and the solvent COM and on the relative orientation
of both particles with respect to the vector joining them. This reduces
the number of Euler angles needed to describe the relative orientation
between the particles from 5 in the general case to only 2 for spherical
solutes. In the same spirit as the treatment of bulk solvent correlations,
we have developed a 1D integral equation theory (1D-IET) approach
to cope with spherical solute dissolved in a general solvent. 1D solute-solvent
correlations are expressed as a function of the distance $r$ and
two Euler angles rather than the 3D position $\bm{r}$ and three Euler
angles. It uses the robust integral equation machinery\citep{blum_invariant_1972,blum_invariant_1972-1,fries_solution_1985,belloni_efficient_2014}
developed for the case of molecular particles since the 1970's to
solve integral equation \ref{eq:rho=00003DExp}. The solute-solvent
OZ equation \ref{eq:gamma} is replaced in practice by the alternate
expression:
\begin{equation}
\gamma_{\text{A}}(r,\bm{\Omega})=\sum_{\text{B}=1}^{N}\rho_{0\text{B}}\iint h_{\text{AB}}(\left\Vert r-r_{2}\right\Vert ,\bm{\Omega},\bm{\Omega}_{2})c_{\text{B}}(r_{2},\bm{\Omega}_{2})dr_{2}d\bm{\Omega}_{2}.\label{eq:gamma_1D}
\end{equation}
The $h_{\text{AB}}$ functions represents the solvent-solvent total
pair correlation function in the bulk, unperturbed, fluid. The
solute-solvent direct correlation function, $c_{\text{B}}$, is related
to the other quantities through
\begin{equation}
\gamma_{\text{B}}=h_{\text{B}}-c_{\text{B}}
\end{equation}
\begin{equation}
h_{\text{B}}=g_{\text{B}}-1=\frac{\rho_{\text{B}}}{\rho_{0\text{B}}}-1.
\end{equation}
Equations \ref{eq:gamma} and \ref{eq:gamma_1D} are strictly equivalent
since the $h_{\text{AB}}$ and $c_{\text{AB}}$ functions are themselves
linked through the solvent-solvent OZ equations. A cycle in the standard
1D IE resolution starts with a guess for $\gamma_{A}$, which allows
to calculate $g_{A}$ from the integral equation \ref{eq:rho=00003DExp}
from which $c_{\text{A}}=g_{\text{A}}-1-\gamma_{\text{A}}$ is deduced.
The solute-solvent projections $c_{\text{A}\mu\nu}^{\ \ mnl}(r)$
are computed in the laboratory frame ($m=\mu=0$ and $n=l$ for spherical
solutes) and transformed into the reciprocal space through 1D Fourier-Hankel
transforms. This allows to express the OZ equation \ref{eq:gamma_1D}
as products between projections in the intermediate inter-molecular
frame. The final $\gamma_{A}$ is obtained through the inverse operations.

It is important to highlight that, contrary to the 3D MDFT implementation,
the charged species do not cause any numerical problem here. Indeed,
the electroneutrality condition in the pure solvent imposes exactly
\begin{equation}
\hat{h}_{++}(0)=\hat{h}_{--}(0)=\hat{h}_{+-}(0)-\frac{1}{n_{+0}}.
\end{equation}
These strict equalities manage without difficulty the analytically known divergence of the $\hat{c}_\text{AB}(q)$ functions at $k=0$. 3D-MDFT and 1D-IET are formally equivalent, except that the 3D-MDFT
implementation deals with orthorhombic periodic boundaries while 1D-IET
implementation assumes a, non-periodic, infinite medium. The direct
comparison of those two numerical approaches will serve as joint validation
of both methods.

After this general presentation of MDFT for a $N$ component mixture
of molecular species we will focus on the particular case of 1:1 electrolyte
aqueous solution. 

\subsection{Aqueous electrolytic solution}

\subsubsection{Three components mixture}

We consider here a 1:1 aqueous electrolyte, \textit{i.e }a mixture
constituted of water, a cationic species and an anionic species. Without
loss of generality, we will focus on 1 M NaCl dissolved in SPC/E water
whose force-field parameters are given in table \ref{tab:Lennard-Jones-parameters-forNaCl}, the temperature is $T=298.15$~K.
\begin{table}
\centering{}%
\begin{tabular}{|c|c|c|c|}
\hline 
species & $\sigma\ (\textrm{Å})$ & $\epsilon\ $(kJ.mol$^{-1}$) & charge ($e$)\tabularnewline
\hline 
\hline 
water (O/H) & 3.165/0.0 & 0.65/0.0  & -0.8476/0.4238\tabularnewline
\hline 
Na$^{+}$ & 2.583 & 0.416 & 1\tabularnewline
\hline 
Cl$^{-}$ & 4.401 & 0.416 & -1\tabularnewline
\hline 
\end{tabular}\caption{Force field parameters for the NaCl aqueous solution\label{tab:Lennard-Jones-parameters-forNaCl}}
\end{table}

In MDFT, each component of the fluid is described by its density.
Since Na$^{+}$ and Cl$^{-}$ are 1 site spherical particles their
densities $n_{+}(\bm{r})$ and $n_{-}(\bm{r})$ solely depend on spatial
coordinates while the water density $\rho_{w}(\bm{r},\bm{\Omega})$
has an additional angular dependancy. Thus, the components of the
functional of equation \ref{eq:F=00003DFid+fExc+Fext} take the following
expression
\begin{equation}
\begin{aligned}F_{\text{id}}[n_{+},n_{-},\rho_{w}] & =k_\text{B}T\int\left(n_{+}(\rr)\ln\left(\frac{n_{+}(\rr)}{n_{\text{0}}(\rr)}\right)-\Delta n_{+}(\rr)\right)d\rr\\
 & +k_\text{B}T\int\left(n_{-}(\rr)\ln\left(\frac{n_{-}(\rr)}{n_{0}(\rr)}\right)-\Delta n_{-}(\rr)\right)d\rr\\
 & +k_\text{B}T\iint\left[\rho_{w}(\bm{r},\bm{\Omega})\ln\left(\frac{\rho_{w}(\bm{r},\bm{\Omega})}{\rho_{0w}}\right)-\Delta\rho_{w}(\bm{r},\bm{\Omega})\right]d\bm{r}d\bm{\Omega}
\end{aligned}
\label{eq:Fidwater}
\end{equation}
 with $\rho_{0w}=n_{0w}/(8\pi^{2})$ where $n_{0w}=0.0326\ \textrm{Å}^{-3}$
is the number density of water and $n_{0+}=n_{0-}=5.94\times10^{-4}\textrm{Å}^{-3}$
are the densities of ions in the bulk mixture. The external part of the
functional reads
\begin{equation}
F_{\text{ext}}[n_{+},n_{-},\rho_{w}]=\int n_{+}(\rr)V_{\text{ext}}^{+}(\rr)d\rr+\int n_{-}(\rr)V_{\text{ext}}^{-}(\rr)d\rr+\iint\rho_{w}(\bm{r},\bm{\Omega})V_{\text{ext}}^{w}(\bm{r},\bm{\Omega})d\bm{r}d\bm{\Omega}.\label{eq:Fext_water}
\end{equation}

Neglecting the bridge functional, the excess term of equation \ref{eq:Fexc}
now is
\begin{align}
F_{\text{exc}}[n_{+},n_{-},\rho_{w}]= & -\frac{1}{2}k_\text{B}T\sum_{y=+,-}\sum_{x=+,-}\iint\Delta n_{x}(\bm{r}_{1})c_{xy}(\left\Vert \bm{r}_{1}-\bm{r}_{2}\right\Vert )\Delta n_{y}(\bm{r}_{2})d\bm{r}_{1}d\bm{r}_{2}\nonumber \\
 & -k_\text{B}T\sum_{x=+,-}\iiint\Delta n_{x}(\bm{r}_{1})c_{xw}(\left\Vert \bm{r}_{1}-\bm{r}_{2}\right\Vert ,\bm{\Omega}_{2})\Delta\rho_{w}(\bm{r}_{2},\bm{\Omega}_{2})d\bm{r}_{1}d\bm{r}_{2}d\bm{\Omega}_{2}\nonumber \\
 & -\frac{1}{2}k_\text{B}T\iiiint\Delta\rho_{w}(\bm{r}_{1},\bm{\Omega}_{1})c_{ww}(\left\Vert \bm{r}_{1}-\bm{r}_{2}\right\Vert ,\bm{\Omega}_{1},\bm{\Omega}_{2})\Delta\rho_{w}(\bm{r}_{2},\bm{\Omega}_{2})d\bm{r}_{1}d\bm{r}_{2}d\bm{\Omega}_{1}d\bm{\Omega}_{2}.\label{eq:Fexc_mixt}
\end{align}

The angular convolution is efficiently performed by developing the
water-water and water-ions convolution products onto a basis of rotational
invariants as in equation \ref{eq:gamma en proj}. Since the ions
are spherical the expansion is complete for $m_{\text{max}}=0$, while
for water the basis set is infinite and needs to be truncated. Choosing
$m_{\text{max}}=4$ proves to be sufficient to reach a good enough
precision. Thanks to the symmetry of the water molecule, the numbers
of independent projections for bulk water-water and ion-water dcf,
$c_{AB\mu\nu;\chi}^{mn}$, are 250 and 9, respectively. The number
of independent projection for the water density, $\rho_{w\mu;\chi}^{m}$
is 55. This corresponds to 225 independent angles for the angular
grid. 

\begin{figure}
\centering{}\includegraphics[width=0.5\textwidth]{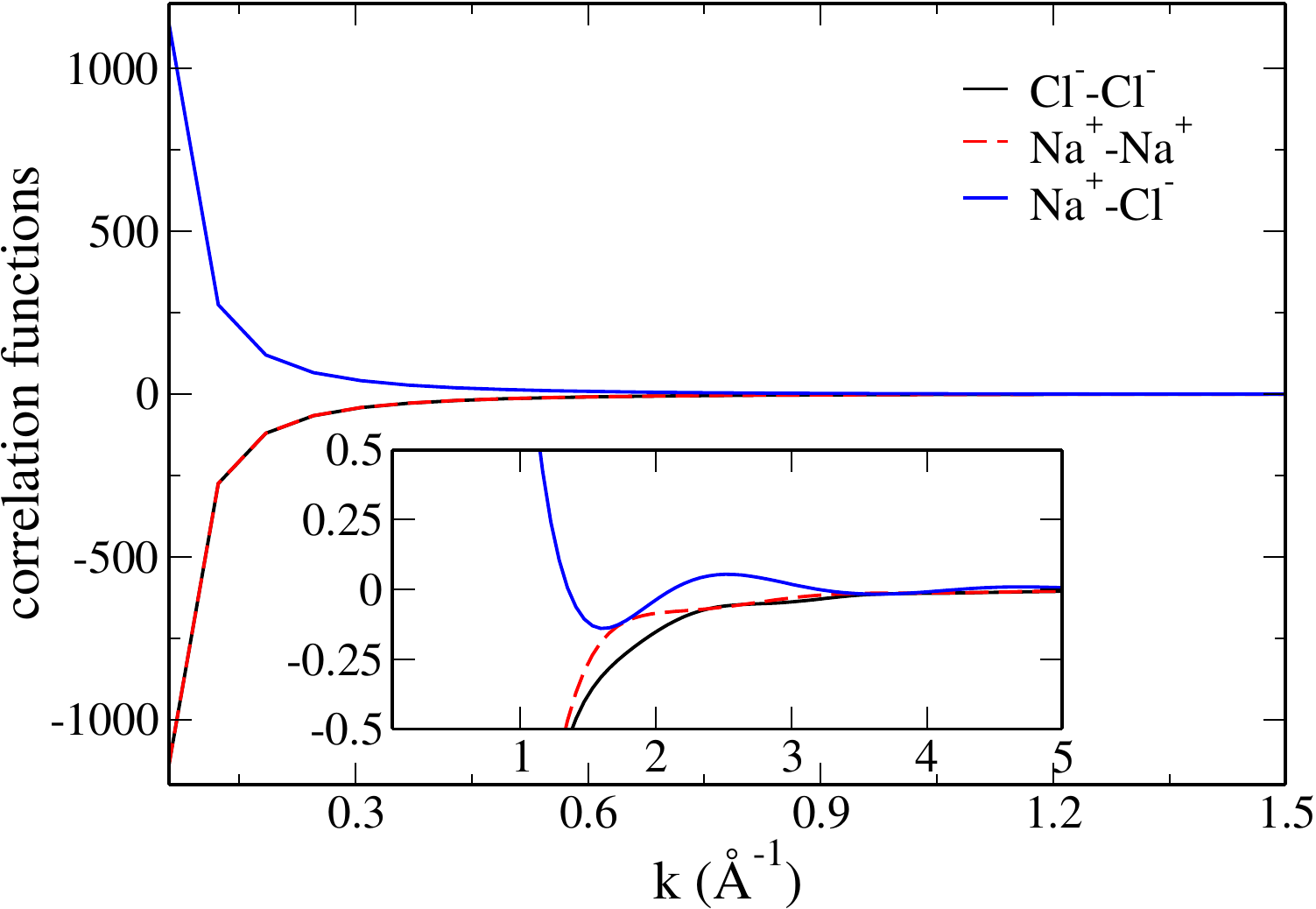}\caption{Ion-ion direct correlation functions, $c_\text{AB}(k)$,  in the SPC/E NaCl (1M) solution.\label{fig:Correlations-functions-electrolyte}}
\end{figure}

The ion-ion dcf employed in the functional of equation \ref{eq:Fexc_mixt}
are depicted in figure \ref{fig:Correlations-functions-electrolyte}.
The water-water, ion-water and water-ion dcf are not shown, but two
points are noteworthy. Firstly, the water-water dcf do not exhibit
significant changes compared to the pure solvent case. Secondly, the
$\hat{c}_{00;0}^{01}(k)$ projection diverges as $-4 \pi  L_Bq\mu i/\sqrt{3}k$ at small
$k$ for ion-water pairs, where $qe$ is the charge of the ion, $\mu e$
is the dipole moment of water and $i$ is the imaginary unit. The
ion-ion dcf diverge asymptotically as $-4\pi L_{B}qq^{\prime}/k^{2}$
at small $k$, where $L_{B}$ is the Bjerrum length, and converge
towards 0 at high values of $k$. In the intermediate region, each
correlations function present some oscillation, as evidenced in the
inset of figure \ref{fig:Correlations-functions-electrolyte}. This
departure from the asymptotic behaviour has two origins; it is a consequence
of the finite size effect arising from the Lennard-Jones interaction
between ion pairs and of the correlations with water. In the Poisson-Boltzmann
case, such oscillation would be absent and the dcf would coincide
with their low $k$ asymptotic behaviour over the entire range of
$k.$ 

\subsubsection{Two components mixture: primitive-like model}

\begin{figure}
\centering{}\includegraphics[width=0.5\textwidth]{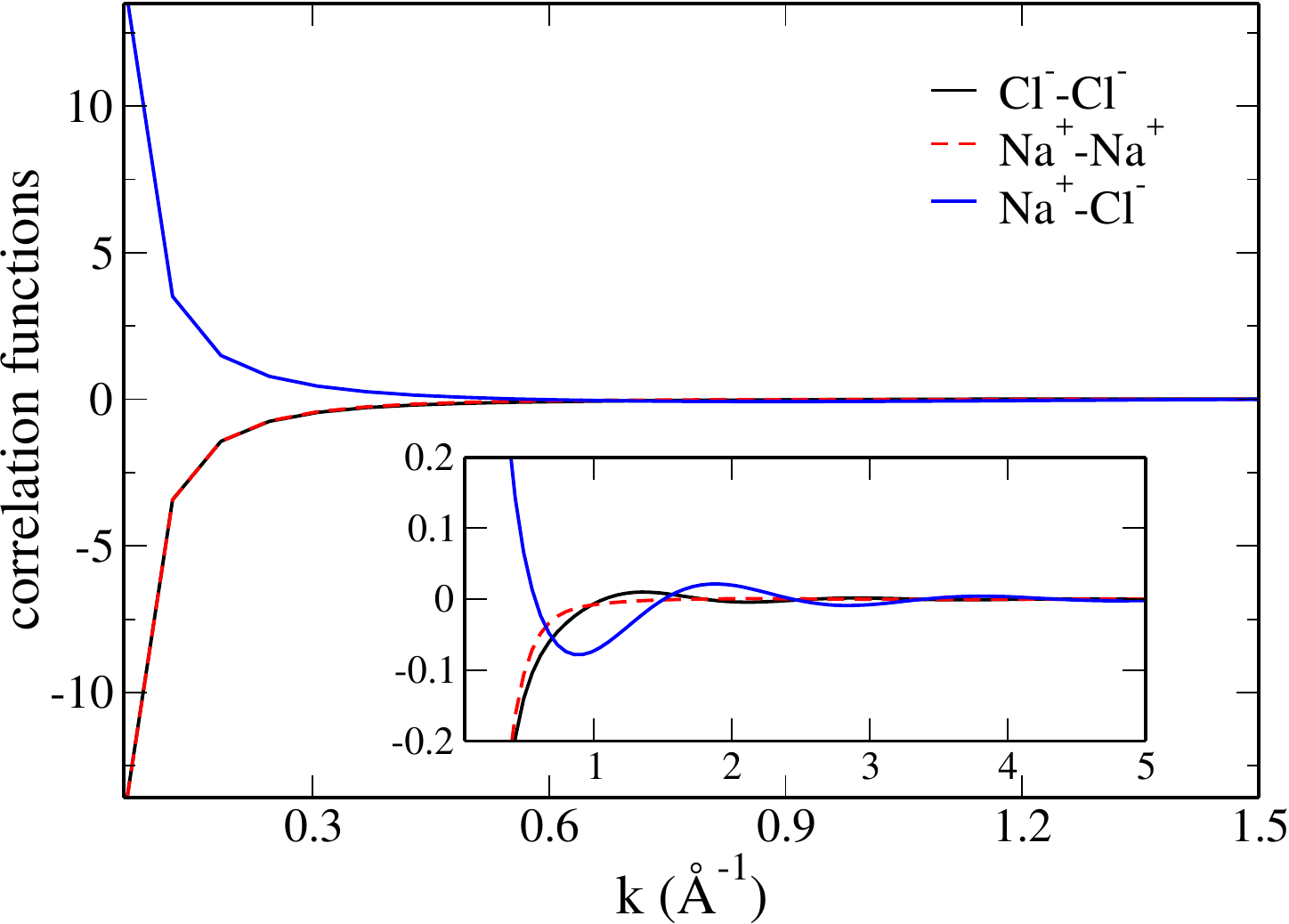}\caption{Ion-ion direct correlation functions, $c_\text{AB}(k)$, in the NaCl (1M) primitive-like model.\label{fig:Correlations-functions-1}}
\end{figure}

If we compare the number of projections for the different dcf used
to describe the ternary NaCl solution, we notice that only 3 are required
to describe the ion-ion terms while 268 involve water. Most of the
computational demand originates from the water-water and water-ion
interactions. An implicit description of water would reduce drastically
the computational cost of the calculation. We thus propose to build
a functional for a two-components primitive-like model of 1M NaCl.
The Lennard-Jones parameters of the ions remain the same as in table
\ref{tab:Lennard-Jones-parameters-forNaCl}, but the electrostatic
interactions are screened by a factor $\epsilon_{r}=78$. The expression
of the functional is obtained by removing all terms involving water
in equations \ref{eq:Fidwater}-\ref{eq:Fexc_mixt}. Is is worth emphasising
that the ion-ion dcfs, which are displayed in figure \ref{fig:Correlations-functions-1},
differ from the one used in the ternary mixture. The most obvious
feature is their amplitudes being reduced by a factor $\epsilon_{r}$
while maintaining an overall similar shape. Noticeable differences
are also present around 2-3 $\textrm{Å}^{-1}$, as evidenced in the
insets of figures \ref{fig:Correlations-functions-electrolyte} and \ref{fig:Correlations-functions-1},
because of modified short-range correlations between the ions without
water.

\section{Results\label{sec:Results}}

\subsection{Primitive Model: test particle insertion}

As a first test case for our primitive model implementation, we focus
on the test particle insertion, i.e the solute is identical to an
ion of the solvent mixture. We use cubic box of $30^{3}\ \textrm{Å}^{3}$
with a grid resolution of 4 points per $\textrm{Å}$. We apply the
usual type-B correction of  Kastenholz \textit{et al}\cite{kastenholz_computation_2006-1} to eliminate contributions arising from
periodic replicas of the inserted particle.

%

\begin{figure}
\centering{}\includegraphics[width=0.5\textwidth]{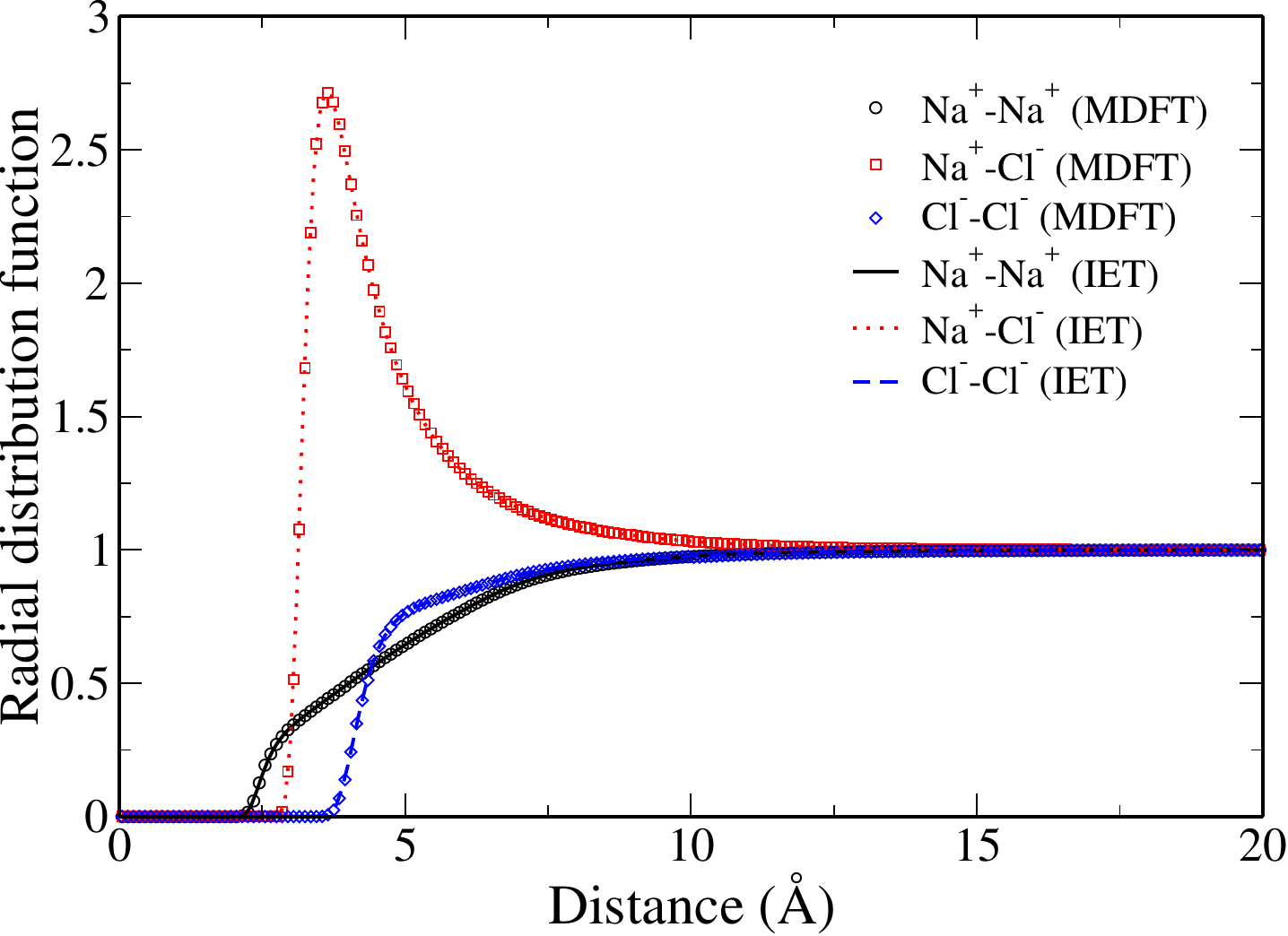}\caption{Radial distributions functions between the test ion and ions of the
primitive solution determined by using IET (lines) and by MDFT (symbols).\label{fig:rdf_insertions_ptcle}}
\end{figure}

In Figure \ref{fig:rdf_insertions_ptcle}, we have compared the radial
distribution functions obtained by solving the 1D IET (lines) and
by 3D MDFT (symbols). The same direct correlation functions (dcf)
were used in both calculations. The agreement is excellent for all
three pairs of ions, demonstrating that the 3D MDFT implementation
reproduces faithfully the results obtained with the 1D IET resolution.

The solvation free energies, which in this particular case equal the
excess chemical potentials of the inserted particles, are reported
in Table \ref{tab:energies_ptcle_insertion}. Once again, the agreement
between MDFT and IET is nearly perfect.

\begin{table}
\centering{}%
\begin{tabular}{|c|c|c|c|}
\hline 
solvent & ion & $\mu_{\text{MDFT}}$ (kJ.mol$^{-1}$) & $\mu_{\text{IET}}\ $(kJ.mol$^{-1}$)\tabularnewline
\hline 
\hline 
primitive & Na$^{+}$ & -1.35 & -1.36\tabularnewline
\hline 
primitive & Cl$^{-}$ & -1.40 & -1.41\tabularnewline
\hline 
\hline 
aqueous & Na$^{+}$ & -390.14 & -390.37\tabularnewline
\hline 
\end{tabular}\caption{Solvation free energies of the inserted ions, as predicted by MDFT
and IET. The 2 first lines corresponds to solvation in the primitive
model while in the last one water is included explicitly as a third
density field \label{tab:energies_ptcle_insertion}}
\end{table}

Again, the main advantage of the 3D MDFT approach over standard IET
is that it can be used to solvate any solutes of arbitrary 3D shape. 

\begin{figure}
\centering{}%
\begin{tabular}{cc}
\includegraphics[width=0.5\textwidth]{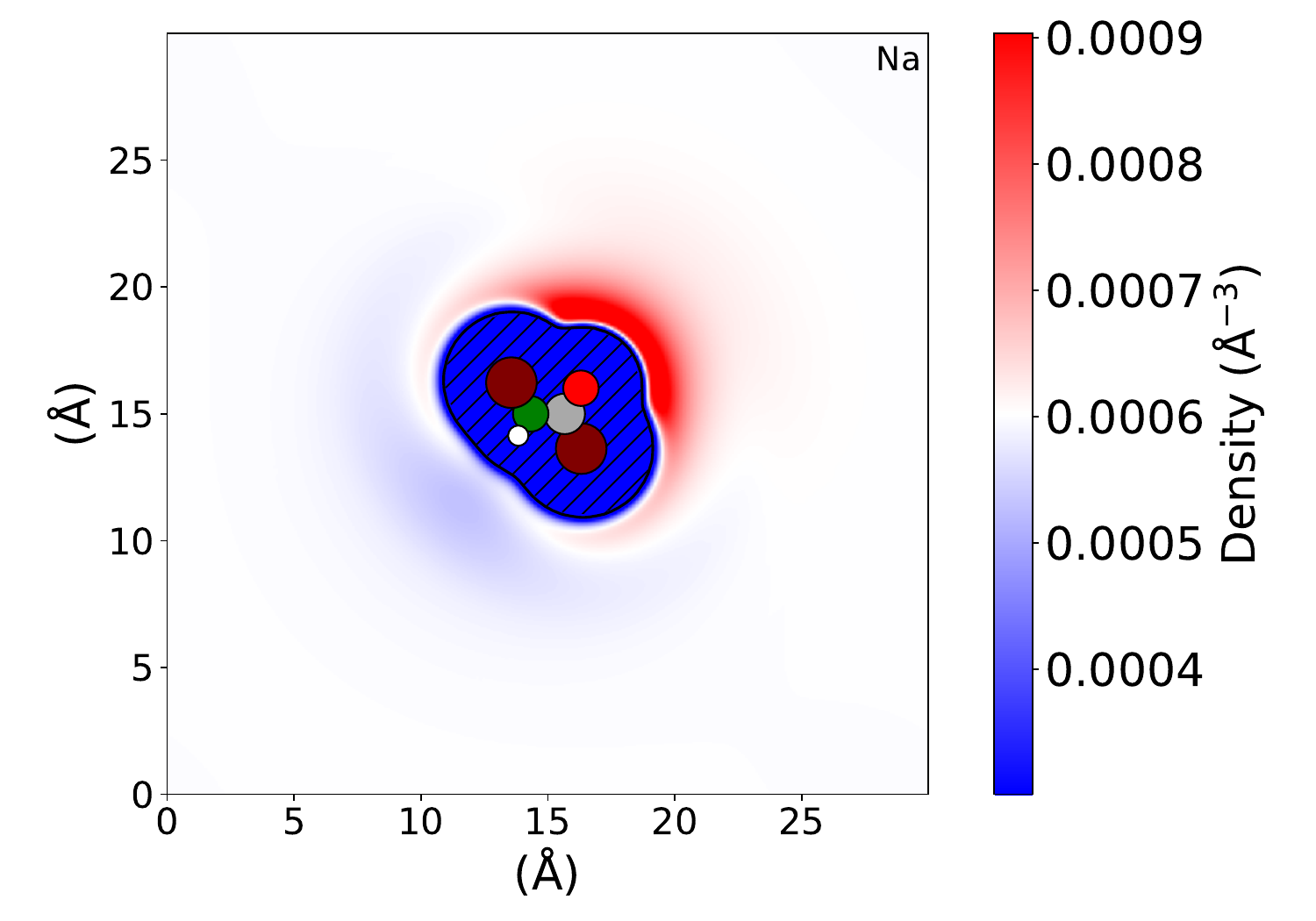} & \includegraphics[width=0.5\textwidth]{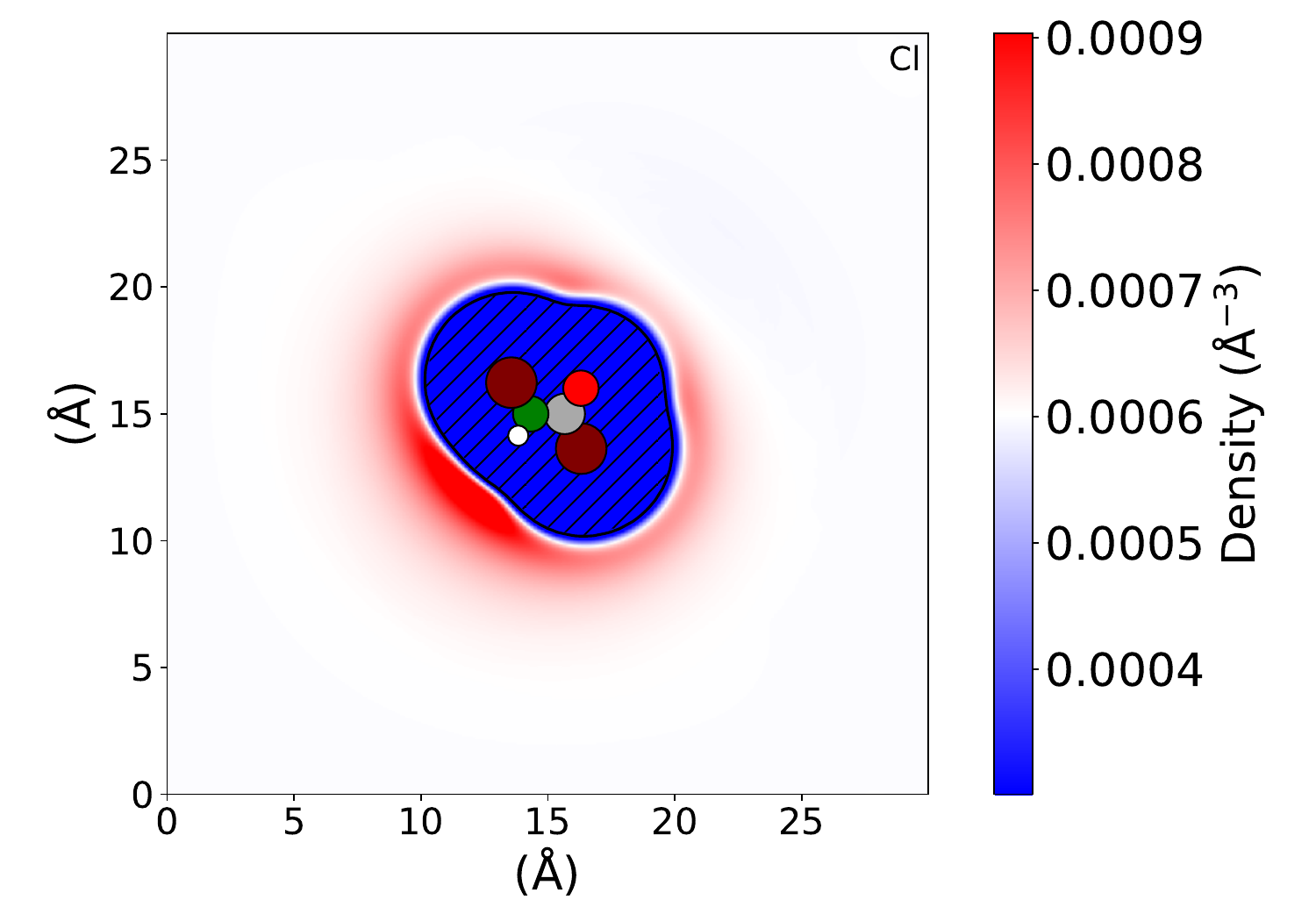}\tabularnewline
\hspace{0.25cm}\includegraphics[width=0.5\textwidth]{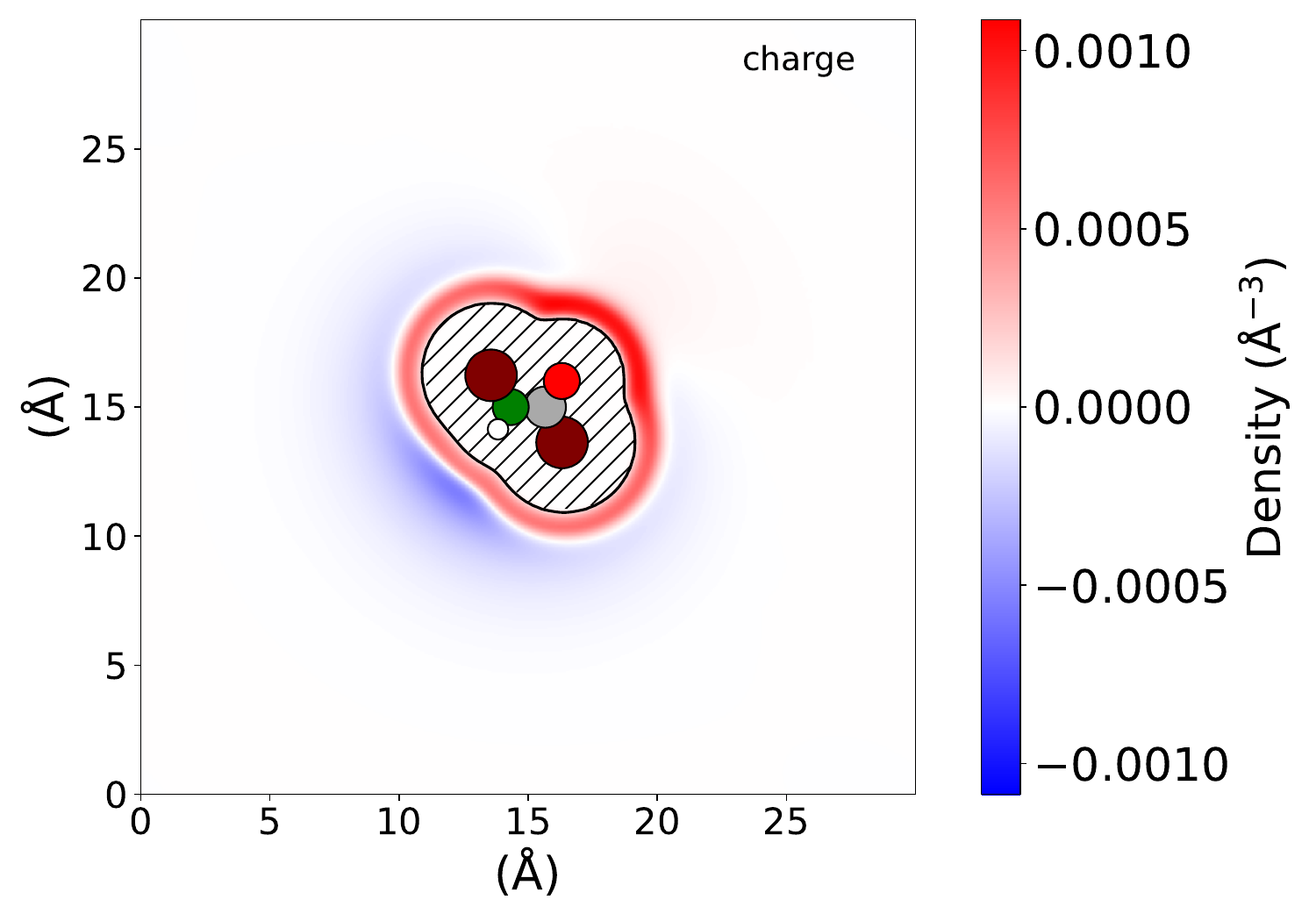} & \includegraphics[viewport=22.20812bp -44.4301bp 372.208bp 298.5699bp,width=0.3\textwidth]{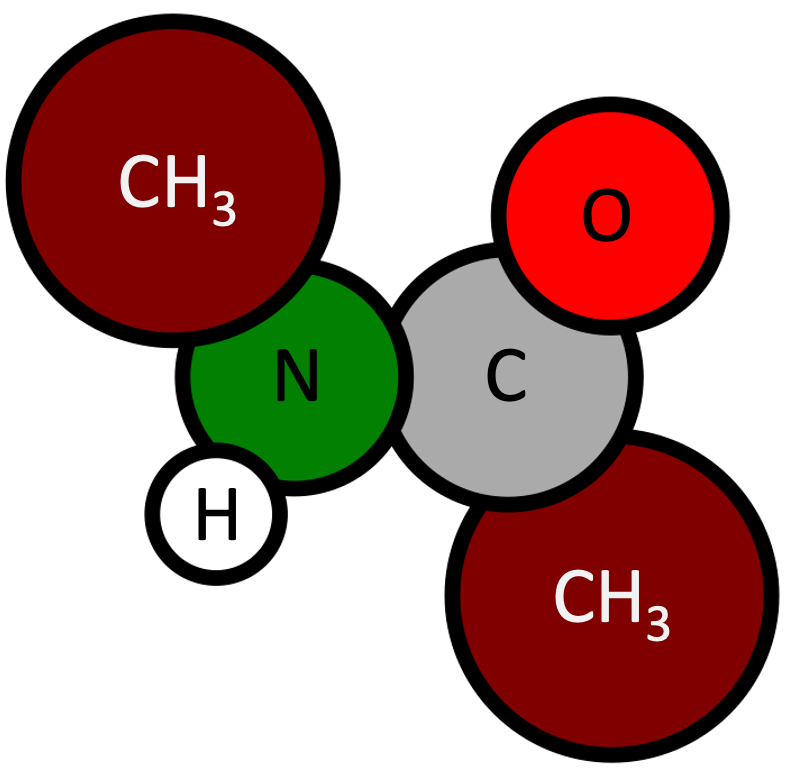}\tabularnewline
\end{tabular}\caption{Slices of density in the plane of the NMA molecule. The top panels
display the sodium density (left) and the chloride density (right).
The bottom panels show the charge density (left), defined as the difference
between the sodium density and the chloride density. The right panel
explains the colour code for the NMA molecule. \label{fig:Slices-NMA-primitve} }
\end{figure}

To illustrate this functionality, we studied the solvation of the
N-methylacetamide (NMA) molecule into the NaCl (1M) primitive-like
model. The force-field parameters (charges and Lennard-Jones) are
available in SI. In figure \ref{fig:Slices-NMA-primitve}, we display
the densities of chloride and sodium in the plane of the NMA molecule.
Our MDFT formulation effectively captures the non trivial arrangement
of the ions around the NMA molecule.

There is a cavity around the solute from which the ions are expelled,
as evidenced by the dashed areas. This cavity is mostly due to finite
size effects caused by the Lennard-Jones interaction, as shown by
the larger cavity in the case of the chloride ion, which has a larger
Lennard-Jones radius than the sodium ion. At the edge of the cavity
begins the first solvation shell of the ions around the NMA molecule. 

For the sodium ions, we observe an accumulation around the oxygen
atom and a depletion around the hydrogen atom, while the first solvation
shell of chloride shows the opposite trend. These trends arise from
the electrostatic interaction: sodium ions are attracted to negatively
charged sites and repelled from positively charged ones, whereas chloride
ions, bearing a negative charge, exhibit the opposite behaviour. Beyond
this first solvation shell, both densities converge toward their bulk
value without further significant oscillations. This is corroborated
by the radial distribution functions (not reported). The solvent charge
density, which is the difference between the sodium density and the
chloride density is also reported in figure \ref{fig:Slices-NMA-primitve}.
Here again, the density follows a similar structure: there is a cavity
around the solute where the density vanishes due to steric exclusion.
This cavity is followed by a first solvation where the charge density
reaches non-zero values. Further from the solute, the charge density
vanishes again due to electroneutrality of the solution. 

Focusing on the solvation shell, there is an excess of sodium ion
in the close vicinity of the cavity followed by an excess of chloride
ion. This indicates that the overall solvation shell is primarily
controlled by steric interactions, i.e. by the difference in cavity
sizes for both ions. The role of the electrostatics is secondary,
causing a more pronounced excess of sodium ions close to the NMA oxygen
which is not followed by an excess of chloride as well as a slight
increase in chloride ion excess near the hydrogen atom. 

This rationalisation of the solvation pattern in term of LJ and electrostatics
contribution has been validated by computing the same quantity for
an NMA molecule with zeroed partial charges, with the resulting density
maps available in SI. The solvent charge density around the neutralised
NMA is similar to the one presented in figure \ref{fig:Slices-NMA-primitve},
but the modulation of the excess of ions in the vicinity of the O
and H atoms of the NMA is no longer present. 

The weak influence of the electrostatic interaction is a limitation
of the primitive model, because all interactions are scaled by $1/\epsilon_{r}$,
including the solute-ions interactions. A more faithful description
of the solvent is thus necessary, which is the topic of the next section.

\subsection{Aqueous electrolyte: test particle insertion}

We now turn to the aqueous electrolyte where water is explicitly represented
by a density field, $\rho_{\text{w}}(\bm{r},\bm{\Omega})$ which depends
on both space coordinates and orientations. Here again, we start by
computing the solvation properties around a sodium ion. We use a $30^{3}\textrm{ Å}^{3}$
calculation box with 3 points per $\textrm{Å}$ and $m_{\text{max}}=0,0,4$
which corresponds to 1, 1 and 225 orientations for sodium, chloride
and water respectively.  We apply the
 type-B correction and   type-C correction of  Kastenholz \textit{et al}\cite{kastenholz_computation_2006-1}.  Type-C correction is due to an improper summation scheme of the solvent polarization due to the periodicity.

\begin{figure}
\begin{centering}
\includegraphics[width=0.5\textwidth]{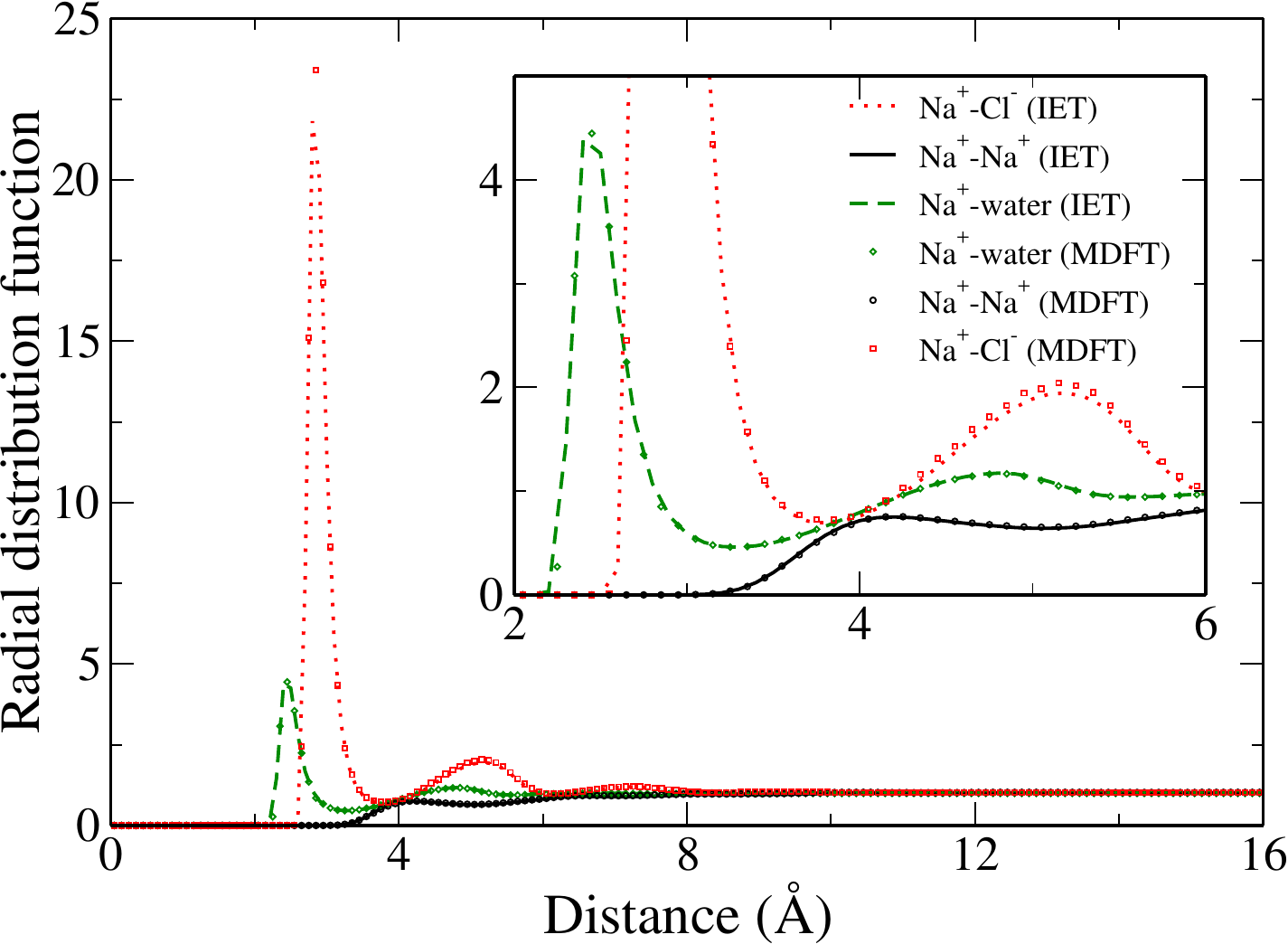}
\par\end{centering}
\centering{}\caption{Radial distributions of aqueous NaCl solution around a Na$^{+}$ cation
computed using IET (lines) and MDFT (symbols)\label{fig:Na_in_waterNaCl}}
\end{figure}

The comparison between the radial distribution
functions obtained with 1D IET and 3D MDFT around a sodium cation
solute is shown in figure \ref{fig:Na_in_waterNaCl}. Again, the agreement
is almost perfect with a slight overestimation of the second pick
of the radial distribution function between Na$^{+}$ and Cl$^{-}$.
Regarding the energetics, the solvation free energy of the sodium
cation predicted by MDFT is $-390.14\ \text{kJ}\cdot\text{mol}^{-1}$
in fairly good agreement with the value of $-390.37\ \text{kJ}\cdot\text{mol}^{-1}$
predicted using 1D-IET. When the solvent is represented explicitly,
the ion-ion radial distribution functions are more structured, exhibiting
several maxima, whereas the primitive-like model shows only a single
maximum. Moreover the first peak of the sodium-chloride radial distribution
function is roughly 10 times higher in the 3 component mixture than
in the primitive model. The central cation is immediately surrounded
by water molecules which is followed by a shell of counterions. It
is worth noting that the sodium cation, despite having the smaller
LJ radius, is expelled quite far from the central cation. This contrasts
with the primitive-like solvent, where the closest contact was between
sodium ions. Clearly, the two solvent descriptions provide two very
different pictures of the solvation properties, even for a simple
spherical cation.

\begin{figure}
\begin{centering}
\begin{tabular}{cc}
\includegraphics[width=0.5\textwidth]{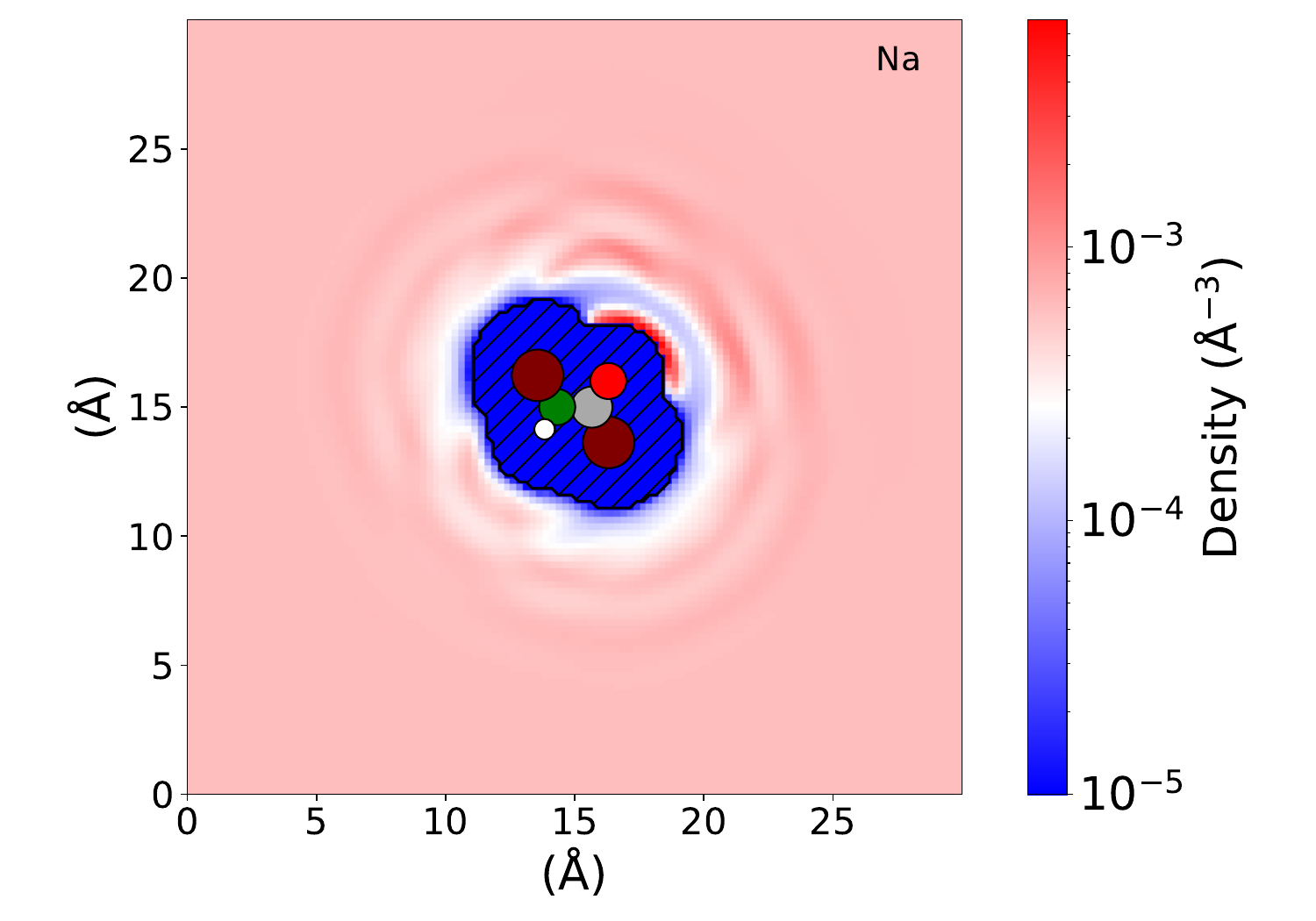} & \includegraphics[width=0.5\textwidth]{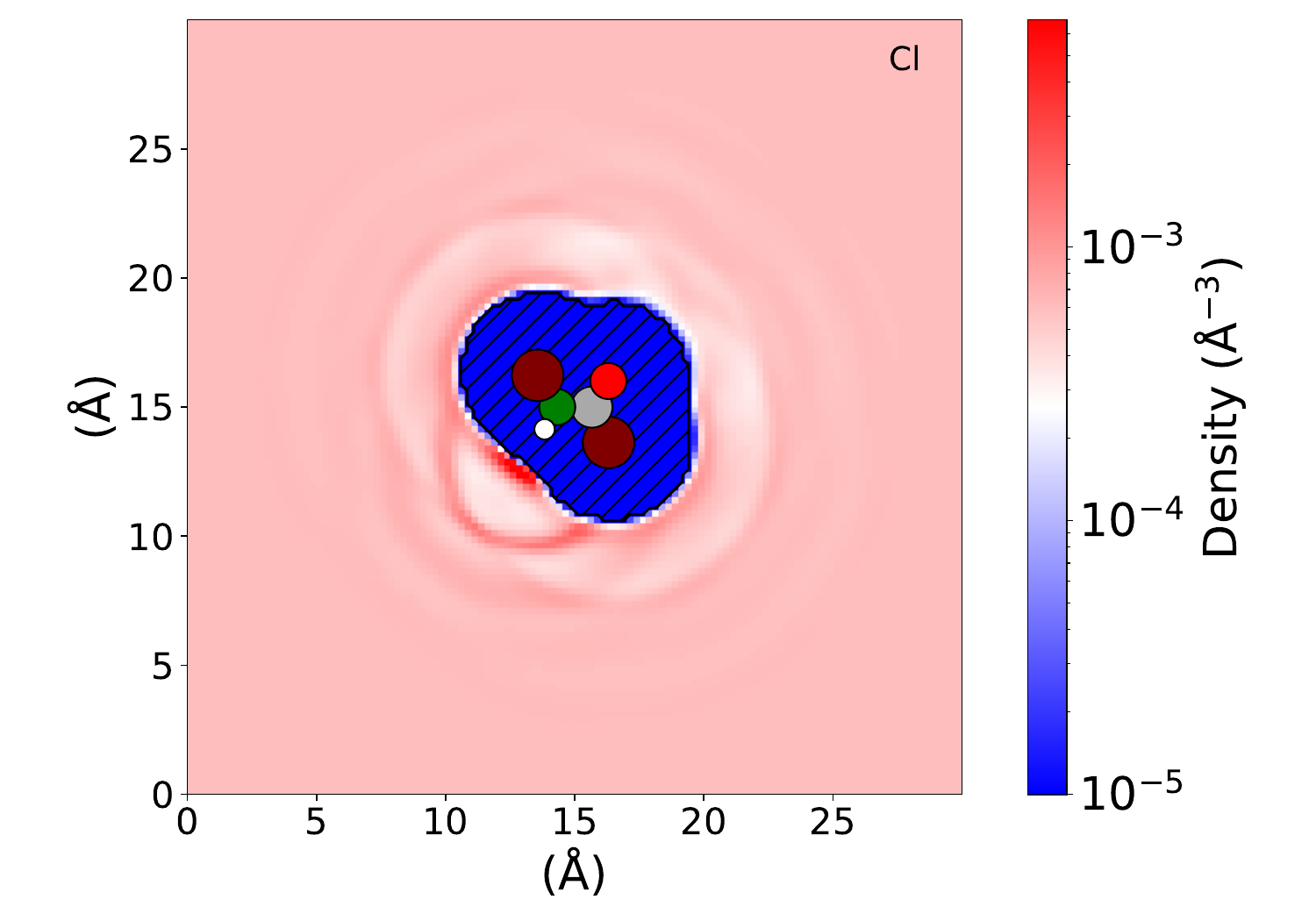}\tabularnewline
\includegraphics[width=0.5\textwidth]{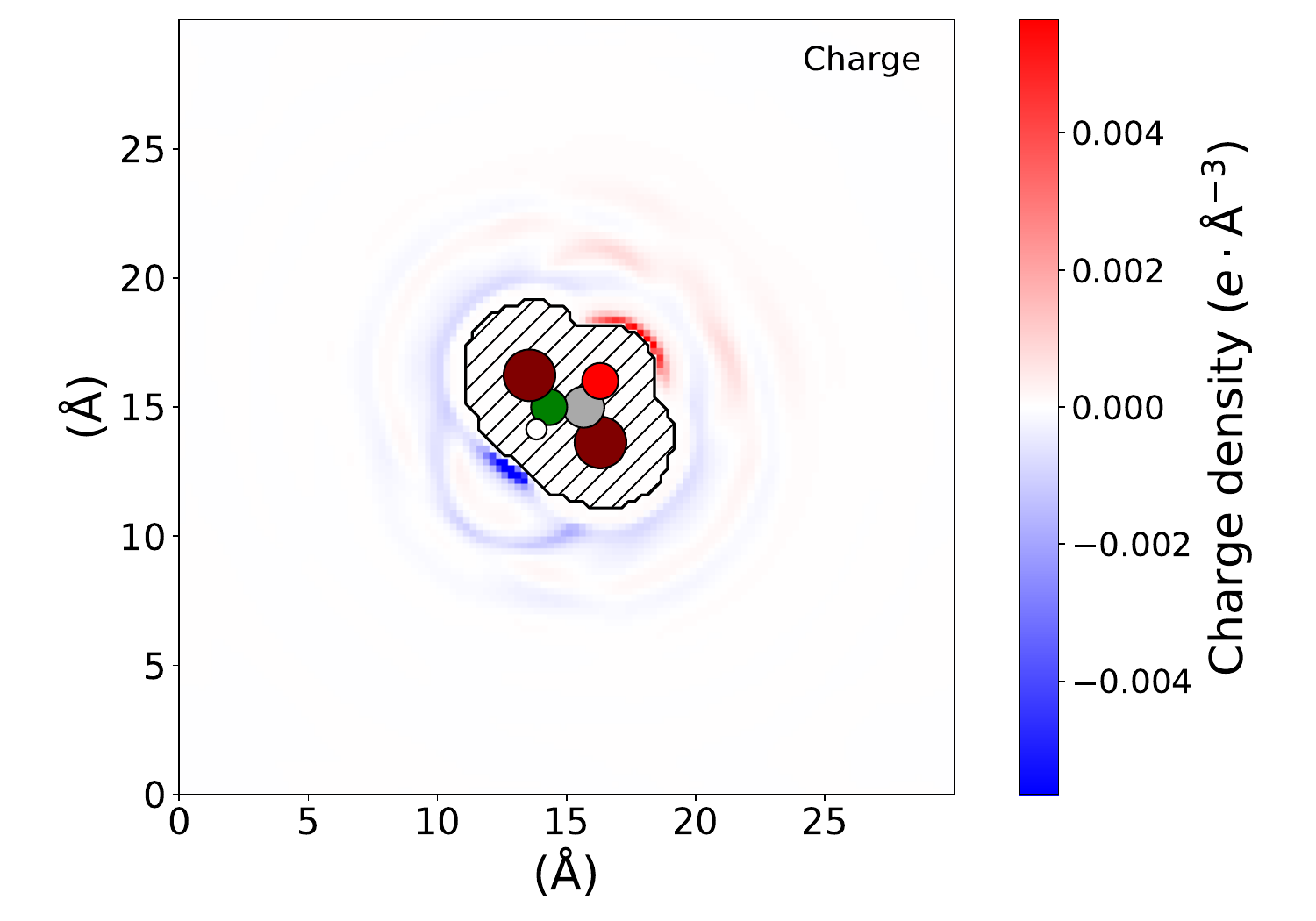} & \includegraphics[width=0.5\textwidth]{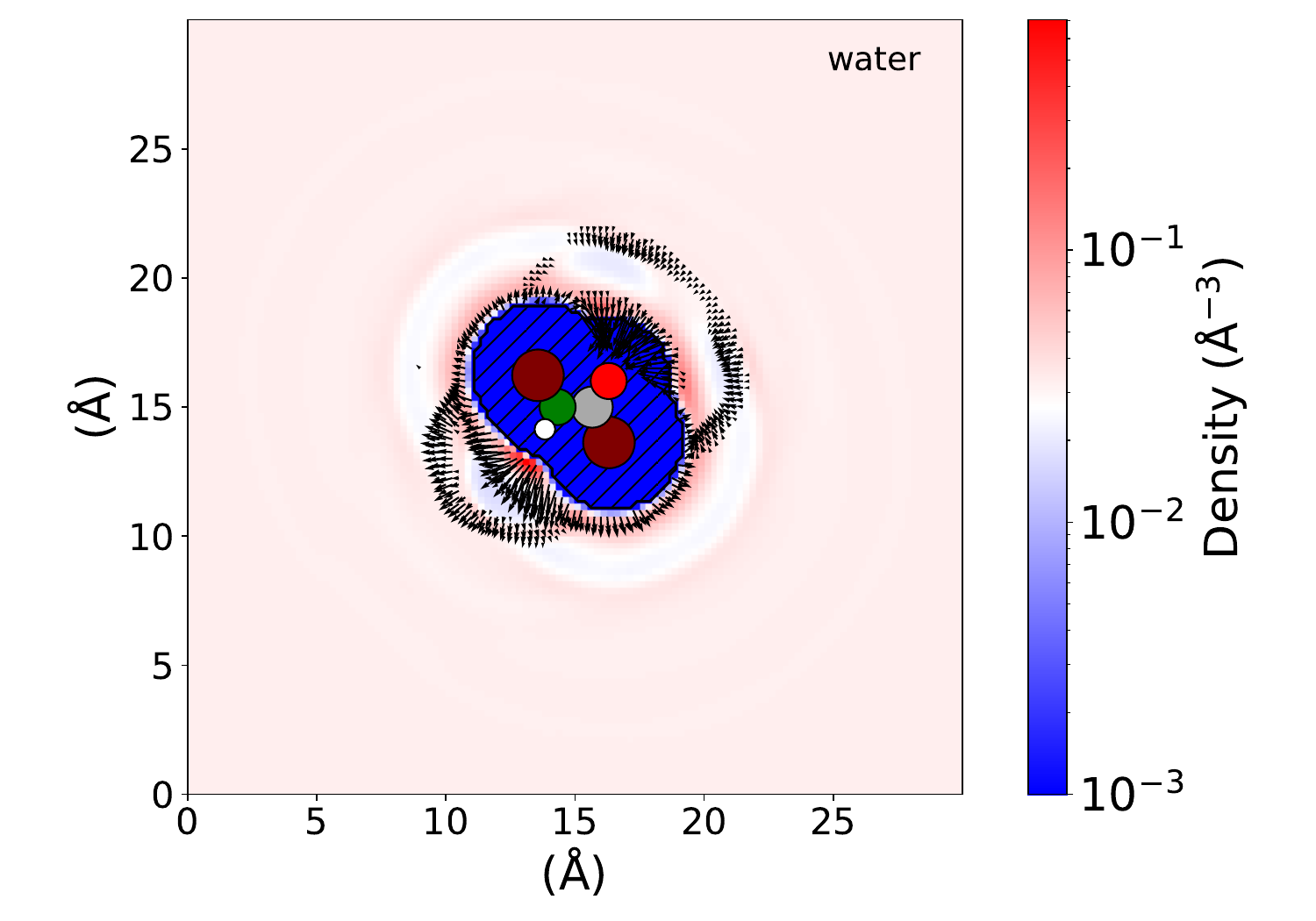}\tabularnewline
\end{tabular}
\par\end{centering}
\centering{}\caption{Slices of density in the plane of the NMA molecule. The top panels
display the sodium density (left) and the chloride density (right).
The bottom panels show the charge density (left), defined as the difference
between the sodium density and the chloride density. The bottom right
panel display the water density. The arrows represents the water polarisation
as defined in equation \ref{eq:pola}.\label{fig:Slices-NMA-electrolye}}
\end{figure}

Again, the primary focus of the present work is to study 3D-shaped
solutes. Figure \ref{fig:Slices-NMA-electrolye} presents density
slices in the plane of the NMA molecule. When focusing on the ions,
it is evident that the solvation differs significantly from the primitive-like
model. For instance, the volume from which the sodium ion is expelled
exhibits a \textquotedbl bump\textquotedbl{} in the vicinity of the
hydrogen atom, which was not present in the primitive-like model.
Both ion densities show oscillations, indicating the presence of several
solvation shells, similar to what was observed in the comparison of
the radial distribution function around the sodium cation.

The charge density around the NMA molecule, defined here as the difference
between the sodium density and the chloride density for consistency
with the primitive like model, exhibits a rather complex structure.
There is an excess of positive charge close to the oxygen atom and
an excess of negative charge close to the nitrogen atom, which is
expected from the electrostatic interactions. However, the successive
layers of positive and negative charges result from the interplay
between the three species and the external potential, making their
a priori prediction challenging. It is also noteworthy that the present
approach properly captures the water structure as evidenced by the
density plot in figure \ref{fig:Slices-NMA-electrolye}. On this plot,
the polarisation which is defined as

\begin{equation}
\bm{P}(\bm{r})=\int\rho_{w}(\bm{r},\bm{\Omega})\bm{\Omega}d\bm{\Omega}\label{eq:pola}
\end{equation}
is represented by arrows. Note that the quantity defined in equation
\ref{eq:pola} is dimensionless and should be multiplied by $\mu$,
the dipole of a water molecule, to recover the dipolar polarization.
For clarity, the preferential orientations are only represented when
$\left\Vert \bm{P}(\bm{r})\right\Vert >\epsilon$. Here, we arbitrarily
choose $\epsilon$ to be one-tenth of the maximum value of $\left\Vert \bm{P}\right\Vert $.
There are 3 regions of high water density around the NMA molecule,
that are found where hydrogen bonding sites with water are expected.
The first one is near the amide hydrogen, with a marked preferential
orientation pointing outward the solute. This indicates a strong interaction
of water molecules with the oxygen pointing towards the N-H bond.
The other two high-density regions are the two lobes close to the
oxygen of the NMA molecule, here the water molecules are preferentially
oriented with their hydrogen atoms pointing towards the oxygen of
NMA. 

\subsection{Conclusions}

In this paper, we presented a generalisation of molecular density
functional theory (MDFT) to mixtures of solvents, with a focus on
the modelling of aqueous electrolytes. While the theoretical extension
of MDFT to mixtures is relatively straightforward, its numerical implementation
proved to be more subtle. Specifically, it was necessary to modify
the relationship between the minimisation variables and the solvent
density to ensure that electroneutrality is maintained throughout
the minimisation process.

The simplest model of an electrolytic solution is the primitive model.
We developed a functional for a similar model consisting of two oppositely
charged Lennard-Jones spheres immersed in a dielectric continuum.
Since the ions are spherical particles, they are represented by a
density $n_{\pm}(\bm{r})$ that does no depend on orientation.

We validated the proposed methodology and its numerical implementation
by computing the solvation properties around a $\text{Na}^{+}$ and
a $\text{Cl}^{-}$ solute in a 1M NaCl primitive model solution. Results
were compared to predictions obtained using integral equation theories
for the same system. We observe nearly perfect agreement in both the
structure and the solvation free energy. The strength of the present
implementation, as compared to integral equation theory, lies in its
ability to model tridimensional solutes. For instance, we were able
to capture in details the solvation structure of $\text{Na}^{+}$
and $\text{Cl}^{-}$ around a N-methylacetamide molecule. 

In the primitive model, where the electrostatic interactions are screened
by a factor 78, the solvation structure is essentially controlled
by the finite size interaction, causing the molecule to be unrealistically
surrounded by sodium cations. Moreover, there is no structuring beyond
the first solvation shell. This limitation of the primitive model
highlights the need for a more refined model of electrolytic solution.
Consequently, we adopted a more realistic three-components model for
aqueous NaCl at 1M where ions are still represented as oppositely
charged spheres but water is now described explicitly using the SPC/E
water model. 

This more advanced description requires knowledge of 6 independent
solvent-solvent direct correlation functions, among which those involving
water have an angular dependancy. Again, we start by comparing the
solvation properties around a $\text{Na}^{+}$ cation predicted by
MDFT and by IET. The agreement between MDFT and IET is excellent for
both structure and free energy. The solvation structure is more complex
than in the primitive model case. For instance, the sodium-chloride
radial distribution function presents a marked peak corresponding
to the first solvation shell. All radial distribution functions exhibit
oscillations revealing the presence of several solvation shells and
long-range ordering. The NMA molecule also presents a more intricate
solvation structure when water is explicitly included, as evidenced
by the oscillations in both water and ion densities and the non trivial
orientational ordering of the water molecules. 

This paper establishes the validity of the functional
approach to model molecular electrolytes. The quality of the predictions
of the presented HNC functional must be assessed in the future by
thorough comparison with explicit simulations and experiments. It
might be necessary to introduce bridge functional, specifically designed
for electrolytes, to complement the HNC functional. Nevertheless,
we believe that MDFT's ability to accurately describe electrolytic
solution at a reasonable computational cost will make it a valuable
tool to address a wide range of systems in which ions play a key role,
such as electrochemical devices and biological systems.

\selectlanguage{french}%
\bibliography{electrolyte}

\selectlanguage{british}%

\end{document}